\documentstyle[preprint,aps]{revtex}
\def\be{\begin{equation}}
\def\ee{\end{equation}}
\def\uln{U_L^{(n)}}
\def\ubln{U_{bL}^{(n)}}
\def\lpn{L^{1/\nu}}
\begin{document}
\draft
\title{\bf
An Extension of Phenomenological Renormalization Method}
\author{M.~Itakura}
\address{Department of Pure and Applied Sciences,\\
University of Tokyo,\\
Meguro-ku, Komaba 3-8-1, Tokyo 153, Japan\\}
\maketitle
\begin{abstract}
We present an extension of the so-called cumulant crossing method
which is used for determination of critical point in Monte Carlo simulations.
The new method uses linear combination of 
several different order-parameter moments and
almost eliminates the systematic deviation of crossing 
points  from the true critical point.
The performance of the method is tested by applying it to the 2D Ising model. 
\end{abstract}
\newpage

The phenomenological renormalization method \cite{BINDER}
has been widely used
in Monte-Carlo simulations of critical phenomena.
It make use of the property of
reduced order-parameter moments
such as
\be
\uln\equiv <M^n>_L/<M^2>_L^{n/2} 
\ee
where $M$ is the order-parameter and
$<\cdots>_L$ denotes the thermal average in
finite system with linear size $L$.
In the vicinity of the critical point,
they obey following scaling form\cite{BINDER}:
\be
\uln \sim f_{n,0}(t \lpn) + L^{-\omega_1}f_{n,1}(t \lpn)
+ L^{-\omega_2}f_{n,2}(t\lpn)+\cdots
\ee
where $t$ is the temperature deviation from the critical point
and $\nu$ is the critical exponent related to 
the correlation length. 
The correction-to-scaling and correction-to-correction-to-scaling exponent etc. \cite{WEGNER}
are denoted by $\omega_i (0<\omega_1<\omega_2<\cdots)$ 
and $f_{n,i}$ denotes associated scaling functions.
Two quantities, $\uln$ and $\ubln$,
intersect at the following temperature:
\be
t\sim C_1 L^{-1/\nu -\omega_1} + C_2 L^{-1/\nu-\omega_2}+\cdots
\label{expand}
\ee
\be
C_i\sim
{(1-b^{-\omega_i})f_{n,i}(0) \over
(b^{1/\nu}-1)f_{n,0}^\prime(0)}
\ee
One can extrapolate the critical point by
plotting the crossing temperature 
versus $L^{-1/\nu-\omega_1}$ .
However, this method requires 
a knowledge about the exponents $\nu$ and $\omega_1$ 
before the estimation of the critical point.\par

We present an extended method 
which reduces the systematic deviation of the crossing points
from the true critical point and does not require any
finite-size scaling.
The key idea is that one can eliminate the 
correction-to-scaling  terms by using linear combination of several 
different  moments.
\be
V_L\equiv
a_4 U^{(4)}_L +
a_6 U^{(6)}_L +\cdots
\ee
It obeys the following scaling form:
\be
V_L\sim
F_0(t \lpn) + L^{-\omega_1}F_1(t \lpn)
+ L^{-\omega_2}F_2(t\lpn)+\cdots
\ee
where
$F_i(x)\equiv a_4 f_{4,i}(x)+a_6 f_{6,i}(x)+\cdots$ 
is the scaling function of $V_L$.
One can adjust the coefficients $a_n$ so that equations 
$F_1(0)=0,F_2(0)=0,\cdots$ holds,
 which leads to $C_1=0,C_2=0,\cdots$ in the equation(\ref{expand}).
For example, 
consider the following simple case:
\be
V_L=a_4 U^{(4)}_L+a_6 U^{(6)}_L
\ee
The
systematic deviation can be reduced
by adjusting the  coefficients $a_n$;
the deviation is approximately proportional to:
\be
{a_4 f_{4,1}(0)+a_6 f_{6,1}(0)\over
a_4 f^\prime_{4,0}(0)+a_6 f^\prime_{6,0}(0)}
L^{-1/\nu-\omega_1}
\ee
Moreover, one can control the direction of the deviation 
by adjusting the coefficients $a_n$
and determine the upper and lower bound
of the critical point.

However,
the estimation of appropriate coefficients $a_n$ is a non-trivial part 
of this method.
We show an example of
the way of estimation in the following.\par
If we plot $U^{(4)}_L$ versus $U^{(6)}_L$ using the values 
at the critical point,
all points are approximately on a straight line.
\be
 (U^{(4)}_L (0), U^{(6)}_L(0))  \sim  (f_{4,0}(0),f_{6,0}(0)) 
 +L^{-\omega_1} (f_{4,1}(0),f_{6,1}(0))
\ee

The slope of the line gives the ratio of two constants 
$f_{4,1}(0)$ and $f_{6,1}(0)$.
The slope at the critical point can be approximated by the value at a
temperature where 
the following condition is satisfied:
\be
{ U^{(4)}_{b L}-U^{(4)}_L \over U^{(6)}_{b L}-U^{(6)}_L}=
{ U^{(4)}_{c L}-U^{(4)}_L \over U^{(6)}_{c L}-U^{(6)}_L}
\ee
where $b \neq c \neq 1$ is some constant.

We have applied this method to Monte Carlo data of the 2D Ising model.
Simulations were done at the critical coupling ($K_c=\log(1+\sqrt{2})$) 
using  Swendsen-Wang  algorithm \cite{SW}
and  $10^5$ observations were made after thermalization.
Positions of crossing points were 
calculated with reweighting techniques \cite{HIST}.
Statistical errors were calculated with jackknife procedure.

Order parameter moments were calculated using cluster size
distribution via the following equations:
\begin{eqnarray}
<M^6>_L & =& 16 <S_6>_L -30 <S_4>_L<S_2>_L +15 <S_2>_L^3 \\
<M^4>_L & =& -2 <S_4>_L +3 <S_2>_L^2 \\
<M^2>_L & =& <S_2>_L \\
 S_n & \equiv & L^{-2n}\sum_j m_j^n
\end{eqnarray}
where $m_j$ is the size of $j$'th cluster in the 
cluster generation step of Swendsen-Wang algorithm.

Fig.1 shows the size-dependence of the deviations of crossing
points in the two cases of
using $U^{(4)}_L$ and $V_L= 5 U^{(4)}_L + U^{(6)}_L$ with $b=1/2$.
One can see a significant reduction of deviation by the use of $V_L$.
The residual deviations of the case of $V_L$ may be the effect of
higher order corrections.
Further extension for eliminating higher
order corrections is straightforward, but it needs observation of 
higher order moments of the order parameter, which is more subjected to
statistical errors.

This work was supported by a Grant-in-Aid for Scientific Research
by the Ministry of Education, Science and Culture.
\newcommand{\cit}[4]{{#1} {\bf #2}, #3 (#4).}
\def\prl{Phys. Rev. Lett.}
\def\pr{Phys. Rev.}
\def\jpsj{J. Phys. Soc. Jpn.}

{\bf Figure Caption }

Fig.1. 
Plot of crossing points in the case of using conventional ($U^{(4)}_L$)
 and new ($V_L$) parameter.

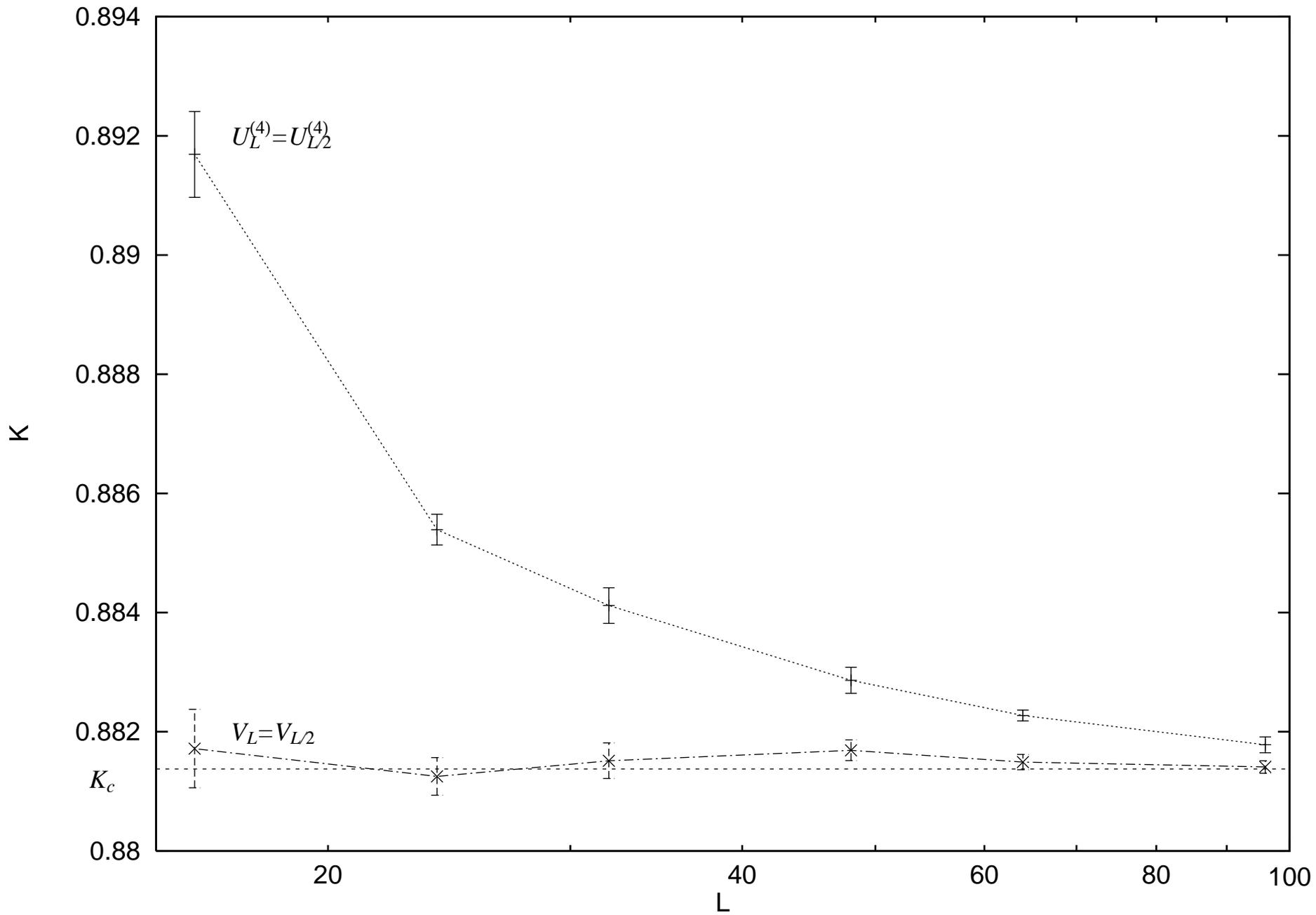
\end{document}